\documentclass[twocolumn,showpreprint,preprintnumbers,amsmath,amssymb]{revtex4}

\usepackage{amsmath}
\usepackage{amssymb}
\usepackage{epsf}
\usepackage{epsfig}
\usepackage{graphicx}
\usepackage{dcolumn}
\usepackage{bm}
\usepackage{psfrag}
\usepackage{textcomp}

\begin{document}

\preprint{}

\title{Separation of the strain and finite size effect on the ferromagnetic properties of La$_{0.5}$Sr$_{0.5}$CoO$_3$ thin films}

\author{Changkun Xie}
\author{J. I. Budnick}
\author{B. O. Wells}%
\affiliation{Department of Physics, University of Connecticut,
Storrs, CT 06269}%
\author{J. C. Woicik}%
\affiliation{National Institute of Standards and Technology,
Gaithersburg, Maryland 20899}%

\date{\today}

\begin{abstract}
The ferromagnetic properties of epitaxial
La$_{0.5}$Sr$_{0.5}$CoO$_3$ thin films have been studied. The
magnetic transition is affected by both strain and finite thickness.
We have used a series of films of different thickness and on
different substrates in order to quantitatively determine the change
in Curie temperature contributed by each effect. The phase diagram
of T$_C$ versus in-plane strain suggests that the ferromagnetic
transition temperature is suppressed by tensile strain and enhanced
by compressive strain. The general method of separating strain and
finite thickness effects should be applicable to any ordering phase
transition in thin films.

\end{abstract}

\maketitle


Thin films of perovskite oxide materials have attracted great
attention lately due to their potential technological applications
based on a variety of appealing physical properties, such as
colossal magnetoresistivity, ferroelectricity, and high-Tc
superconductivity.  The properties of films differ from the
corresponding bulk typically due to a combination of three factors.
Firstly, defect levels are often higher in films. Oxygen deficiency
is the most common defect and will typically suppress the transition
temperature due to the decrease of doped hole density or the
destruction of metal-oxygen hopping pathways.\cite{Dho03} Secondly,
finite size effects may be important. For example, the Curie
temperature (T$_C$) for a ferromagnetic thin film will be reduced
when the spin-spin correlation length exceeds the film thickness.
The thickness-dependent Curie temperature has been most carefully
studied in simple metallic films of Fe, Co, Ni and
Gd.~\cite{js95,Rz01} A similar scaling effect has also been found in
ferroelectric materials.\cite{ca98} Thirdly, strain incorporated
into films due to effects such as a lattice mismatch with the
substrate may also alter the phase transition through changes in
fundamental interactions that depend upon atomic spacing. Strain in
thin films is often thought of as analogous to that induced in
high-pressure experiments on bulk materials. However, a much larger
strain can be achieved in films than that in bulk and the strain in
films is usually biaxial rather than hydrostatic or uniaxial as in
most bulk pressure experiments. The induced strain can modify the
lattice structure, the critical temperature for phase transition,
and sometimes the nature of the phases present
themselves.~\cite{qy04,Lu00,xj05}

La$_{0.5}$Sr$_{0.5}$CoO$_3$(LSCO) is a highly doped ferromagnetic
oxide material with perovskite structure and has desirable
properties of high electrical conductivity and large
magnetoresistance.~\cite{gb95,by97} Thin film LSCO is a candidate
for applications such as electrodes for fuel cells, ferroelectric
memory and spin valve devices. The first concern noted above, the
oxygen content in LSCO films, can be controlled by carefully
optimizing growth conditions so that stoichiometry of oxygen can be
maintained.~\cite{am00} However, the finite size and strain effects
are intrinsic to the film, and it is not trivial to separate the
influence of these two effects on a particular film. Recently, Fuchs
et al ~\cite{Df05} and  Andres et al ~\cite{de03} have reported the
finite thickness effect for the perovskite oxide films,
La$_{0.7}$Sr$_{0.3}$CoO$_3$  and La$_{0.7}$Ca$_{0.3}$MnO$_3$,
respectively. However, the possible effects of strain were
discounted in the analysis. Conversely, studies concerned with the
effect of strain have neglected finite size scaling.~\cite{si00}
Clearly it is necessary to consider both possible effects on an
equal footing and it would be desirable to assign a particular
amount of change in transition temperature to a particular cause. In
this paper, we report a systematic study of the Curie temperature of
a series of ferromagnetic LSCO films with different thickness on a
variety of substrates. This allows us to separate the finite
thickness effect from the direct effect of strain and thus construct
a T$_C$ versus strain phase diagram.

LSCO films were epitaxially grown by a pulsed laser deposition
technique from a stoichiometric target. Deposition was carried out
using a 248 nm KrF excimer laser at 2 Hz, with a substrate
temperature of 750 \textcelsius, in an atmosphere of 200 mTorr
O$_2$. After growth the films were slowly cooled in 300 Torr O$_2$.
The fact that our thickest films with little strain had a T$_C$ very
close to the bulk value ($\approx$ 250 K) is an indication that this
procedure produces film with near stoichiometric oxygen. A set of
films with thickness varying from 11 nm to 126 nm were grown on
LaAlO$_3$ (LAO) and SrTiO$_3$ (STO). For a thickness of 22 nm, LSCO
films were also grown on (LaAlO$_3$)$_{0.3}$(Sr$_2$LaTaO$_6$)$_7$
(LSAT) and SrLaAlO$_4$ (SLAO) substrates.

Structural characterization of the films was carried out using a
conventional powder x-ray diffractometer and a four circle
diffractometer with a general area detector Diffraction System
(GADDS) at room temperature. The only peaks observed in the normal
$\theta$-2$\theta$ scans are from the (00\emph{l}) type peaks
indicating the absence of impurity phase or grains with other
orientations. The in-plane and out-of-plane lattice parameters were
determined by normal and off-axis diffraction. The field-cooled DC
magnetic properties were measured using a superconducting quantum
interference device (SQUID) magnetometer in a range of 5 K to 320 K
at a magnetic field of 100 G with the applied field direction
parallel to the sample surface. Since the transitions were slightly
rounded, we determined $T_C$ by extrapolating the linear portion of
the magnetization curve just below the transition. $T_C$ is defined
as the point where this line intercepts the paramagnetic background.


The in-plane and out-of-plane lattice parameters for films of LSCO
with various thickness on LAO and STO substrates are shown in Fig.
1. These lattice constants determine the strain state of the films.
Bulk LSCO is a distorted perovskite with a cubic lattice parameter
of 0.3834 nm, shown as a dashed line in the figure. On both
substrates, the thinnest films appear to be coherent in plane with
the substrates. The thinnest film on STO is under a large tensile
strain and undergoes a monotonic relaxation as the film thickness
increases. The films on LAO show somewhat different behavior. The
thinner films are under compressive strain, but thicker films show a
small tensile strain similar to the thicker films on STO. This
appears to indicate that the misfit strain at higher temperature is
completely relaxed in these thicker films and the remaining strain
results from the difference in the thermal expansion coefficients
between the films and the substrates.\cite{hg04,pc06}

\begin{figure}
\includegraphics[scale=0.8]{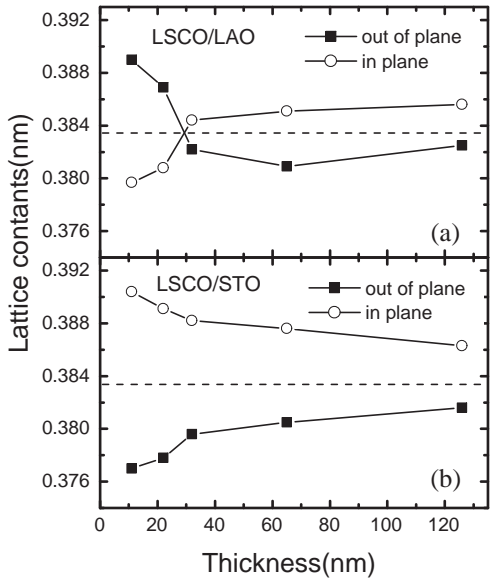}
\caption{\label{fig:fig1.eps}Film thickness dependence of the
measured in-plane (circles) and out-of-plane (squares) lattice
parameters of epitaxial LSCO films grown on (a) (001) LAO and (b)
(001) STO substrates. The bulk parameters of the LSCO target are
indicated by the horizontal dashed line.The size of the error bars
is smaller than that of the symbols.}
\end{figure}

Fig. 2 shows the Curie transition temperature vs film thickness for
LSCO on LAO and STO, respectively. On both substrates, T$_C$ is
substantially reduced for the thinnest films, gradually increases
with increasing thickness, and eventually reaches a value near that
of the bulk. However, the thin films on LAO  show a significantly
higher Curie temperature than those on STO, most markedly in the
thinner regions. As seen from Fig. 1, below 30 nm, the films on STO
are under tensile strain while those on LAO are compressed. Since
the finite size effect ought to be independent of choice of
substrate, it appears that the phase transition temperature is
sensitive to the strain induced by the substrates as well as finite
size.

\begin{figure}
\includegraphics[scale=0.32]{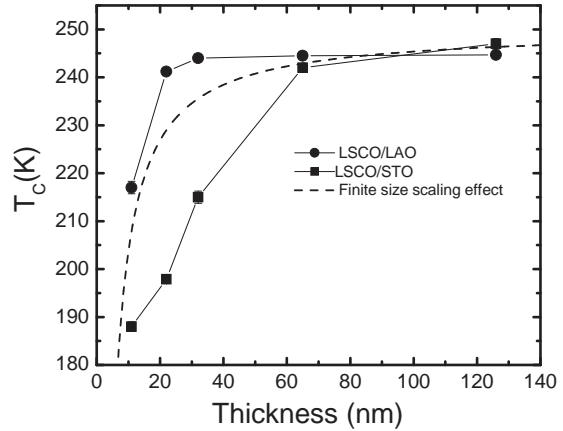}
\caption{\label{fig:fig2.eps}Film thickness dependence of $T_C$ for
epitaxial LSCO thin films grown on LAO and STO. The dashed line is
our calculation for $T_C$ based on finite size scaling alone as
described in the text. }
\end{figure}

It is well known that the finite size effect will suppress the Curie
temperature when the film thickness is below a certain critical
value. According to critical behavior scaling theory for a magnetic
phase transition, \cite{Rz01}

\begin{equation}
[T_c(\infty)-T_c(d)]/T_c(\infty)=(c/d)^{\lambda}
\end{equation}

where $T_c(\infty)$ is the ferromagnetic Curie temperature in the
bulk limit, $d$ is the film thickness, $\lambda$ is the critical
shift exponent whose value reflects the universality class of the
phase transition, and $c$ is a constant related to the spin-spin
correlation length. To produce a sample that is only affected by the
finite size effect one would need a substrate with the same lattice
constants and same thermal expansion as the LSCO films, which is not
possible. We have followed a converse strategy. Thin films with the
same thickness are grown on a variety of substrates with different
lattice constants, so that each film will have the same finite size
scaling effect but different strain. We have chosen to focus on
films of thickness 22 nm on four different substrates. These are
LAO(0.3790 nm), SLAO(0.3756 nm), LSAT(0.3868 nm) and STO(0.3905 nm).
Films on LAO and SLAO are compressed while those on LSAT and STO are
under tension. The phase diagram of the ferromagnetic transition
temperature T$_C$ with in-plane strain for these thin films is shown
in Fig. 3. The Curie temperature appears to be linearly dependent on
the in-plane strain. The interpolation of this plot to the point
with zero strain provides an estimate of the reduction of the Curie
temperature by finite size scaling alone for a thickness of 22 nm.

We find that the Curie temperature for a film without strain at a
thickness of 22 nm would be about 229 K. This allows us to construct
a quantitative prediction of T$_C$ versus thickness due to finite
size scaling. According to the theory,\cite{Rz01} the critical shift
component is expected to be $\lambda$=1 for the mean field
approximation, which is a reasonable description for an itinerant or
double-exchange based ferromagnet.~\cite{me72} Similarly, de Andres
~\cite{de03} has reported that for manganite films the reduction of
T$_C$ is well fit with $\lambda$=1. Thus taking $\lambda$=1 and
T$_C(\infty) \approx$ 250 K,~\cite{sm96} the bulk limit of LSCO,
leaves a single fitting parameter, $c$, for the finite size effect
equation(1). We fit $c$ in the equation to make $T_C$ at 22 nm
thickness equal to 229 K. This gives a value for $c$ of 1.848 nm,
which corresponds to a spin-spin correlation length of about 9
lattice constants according to a model by Zhang et al., similar to
Gd films.\cite{Rz01} The resulting curve is shown as a dashed line
in in Fig. 2. Deviation of the actual Curie Temperature from this
line for a specific film should be primarily due to the effects of
strain on fundamental material parameters.

\begin{figure}
\includegraphics[scale=0.8]{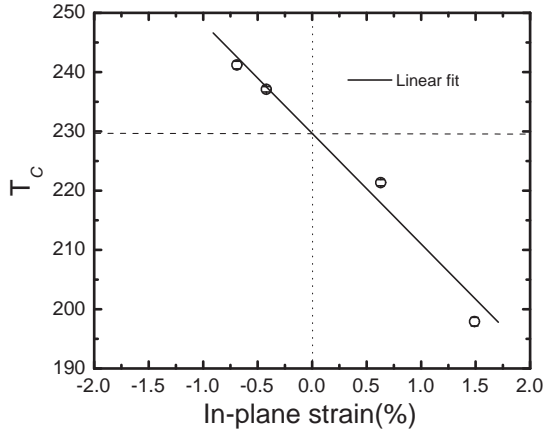}
\caption{\label{fig:fig3.eps}Phase diagram of Ferromagnetic
transition temperature as a function of in-plane strain for films of
thickness 22nm. The interpolation to zero strain indicates a T$_C$
of about 229 K, indicated as the horizontal dashed line. }
\end{figure}

\begin{figure}
\includegraphics[scale=0.8]{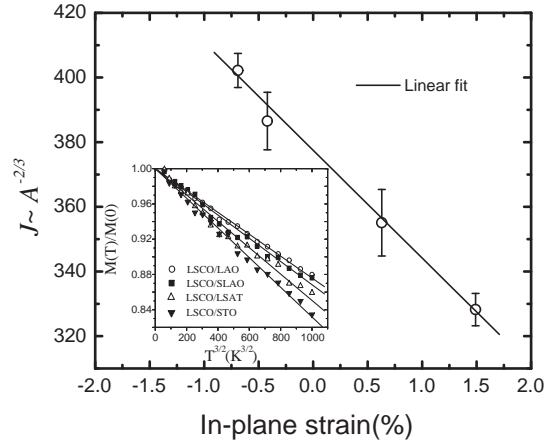}
\caption{\label{fig:fig4.eps}Exchange interaction $J (\sim
A^{-2/3})$ vs. in-plane strain. The inset shows the reduced
magnetization $M(T) /M(0)$ as a function of $T^{3/2}$ with an
applied magnetic field 100 G for the same films as in Fig. 3. We fit
the data according to Bloch's law to derive the spin wave parameter
$A$ and thus $J$.}
\end{figure}

An examination of Fig. 2 and Fig. 3, along with the discussion
above, suggests that in-plane tension suppresses the Curie
Temperature while in-plane compression enhances it. If this is true,
then we expect to see directly a change in the exchange coupling
energy, $J$, that matches the changes in $T_C$ seen in Fig. 3. The
exchange energy can be derived from Bloch's law.~\cite{fh64,bs86}
The variation of the spontaneous magnetization $M(T)$ with the
temperature $T$ at low temperature limit is determined by
$M(T)=M_0(1-AT^{3/2})$ where $M_0$ is the spontaneous magnetization
at 0 K and $A$ is a constant, called spin wave parameter. In this
model, the exchange interaction $J$ is proportional to $A^{-2/3}$.
The reduced magnetization $M(T)/M(0)$ vs $T^{3/2}$ in the low
temperature region is shown in the inset of Fig. 4. The data are fit
to Bloch's law, and the spin wave parameter $A$ is derived. The
exchange interaction, as indicated by $A^{-2/3}$, is plotted in
Fig.4. There is a strong dependence on in-plane strain, very similar
to the linear behavior of the Curie Temperature shown in Fig. 3.
According to mean field theory, $T_C$ is proportional to J in a bulk
magnet.~\cite{vb72} Thus this supports the idea that the primary
factor causing the change in $T_C$ for the films of thickness 22 nm
is a change in J caused by strain. It is precisely these changes in
J that can reveal insights into the origin of magnetism in this
compound. A leading theory for the magnetism in LSCO is the double
exchange mechanism.\cite{pg60} This model relies upon Co-O-Co
electron hopping so that a strong dependence on bond length is
expected. However a proper evaluation of the theory requires an in
depth analysis of the local atomic distortions resulting from
strain.\cite{jw97} In a future report on our EXAFS
results,\cite{tobe} we will examine whether the double exchange
mechanism quantitatively predicts the strain dependence we have
measured.

In summary, the Curie temperature of ferromagnetic
La$_{0.5}$Sr$_{0.5}$CoO$_3$ thin films was studied as a function of
film thickness and strain. Both finite-size and strain affect the
ferromagnetic properties and $T_C$. By studying films with a variety
of thickness, grown on a variety of substrates, we were able to
quantitatively assign the change in $T_C$ to the two contributing
factors. An analysis of magnetization data to give the spin
interaction energy appears to confirm that this assignment is
reasonable. Ferromagnetism is suppressed by tension and enhanced by
compression.

This work is supported through NSF DMR-0239667. BOW acknowledges
support from the Cottrell Scholar program of the Research
Corporation.


\begin{thebibliography}{23}



\bibitem{Dho03} J. Dho, N. H. Hura, I. S. Kim and Y. K. Park, J. Appl. Phys. {\bf 94}, 7670 (2003).

\bibitem{js95} J. S. Jiang, D. Davidovic, D. H. Reich and C. L. Chien, Phys. Rev. Lett. {\bf 74}, 314 (1995).

\bibitem{Rz01} R. Zhang and R. F. Willis, Phys. Rev. Lett. {\bf 86}, 2665 (2001).

\bibitem{ca98} C. A. Randall, D. E. Mccauley and D.P.Cann, Ferroelectrics {\bf 206}, 325 (1998).

\bibitem{qy04} Z. Q. Yang, R. Hendrikx, J. Aarts, Y. L. Qin and H. W. Zandbergen,Phys. Rev. B {\bf 70}, 174111
(2004).

\bibitem{Lu00} Y. Lu, J. Klein, C. Hofener, B. Wiedenhorst, J.B.Philipp, F.Herbstritt, A.Marx, L.Alff and R.Gross,
Phys. Rev. B {\bf 62}, 15806 (2000).


\bibitem{xj05} X. J. Chen, H. -U. Habermeier, H. Zhang, G. Gu, M. Varela, J. Santamaria and C. C. Almasan,
Phys. Rev. B {\bf 72}, 174403 (2005).

\bibitem{gb95} G. Brice\"{n}o, H. Chang, X. Sun, P. G. Schultz,  X. -D. Xiang, Science {\bf 270}, 273 (1995).

\bibitem{by97} B.Yang, S.Aggarwal, A.M.Dhote, T.K.Song, R.Ramesh and J.S.Lee, Appl. Phys. Lett {\bf 71}, 356 (1997).

\bibitem{am00} A. R. Moodenbaugh, B. Nielsen, S. Sambasivan, D. A. Fischer, T. Friessnegg, S. Aggarwal, R. Ramesh and R. L. Pfeffer,
Phys. Rev. B {\bf 61}, 5666 (2000).

\bibitem{Df05} D. Fuchs, T. Schwarz, O. Moris, P. Schweiss and R. Schneider, Phys. Rev. B {\bf 71}, 92406 (2005).

\bibitem{de03} A. de Andres, J. Rubio, G. Castro, S. Taboada, J. L. Martnez and J. M. Colino,
Appl. Phys. Lett. {\bf 83}, 713 (2003).

\bibitem{si00} S. I. Khartsev, P. Johnsson, and A. M. Grishina,
J. Appl. Phys. {\bf 87}, 2394 (2000).

\bibitem{hg04} F. S. Aguirre-Tostado, A. Herrera-G\'{o}mez, J. C. Woicik, R. Droopad, Z. Yu, D. G. Schlom,P. Zschack, E. Karapetrova, P. Pianetta, and C. S. Hellberg,
Phys. Rev. B, {\bf 70}, 201403(R) (2004)

\bibitem{pc06} I. B. Misirlioglu and S. P. Alpay,Institute of Materials Science, University of Connecticut, Private
communication. They have calculated strain due to thermal expansion
of LSCO films on STO and LAO substrates to be consistent with our
data.

\bibitem{me72} M. E. Fisher and M. N. Barber,Phys. Rev. Lett. {\bf 28}, 1516 (1972).

\bibitem{sm96} S. Mukherjee, R. Ranganathan, P.S.Anilkumar and P.A.Joy,
Phys. Rev. B {\bf 54}, 9267 (1996).

\bibitem{fh64} F. Holtzberg, T. R. McGuire, S. Methfessel and J. C. Suits, J. Appl. Phys. {\bf 35}, 1033 (1964).

\bibitem{bs86} B. S. Kang, J. -S. Lee, L. Stan, L. Civale, R. F. DePaula, P. N. Arendt and Q. X. Jia, Appl. Phys. Lett. {\bf 86}, 072511 (2005).

\bibitem{vb72} U. von Barth and L. Hedin, J. Phys. C: Solid State Phys. {\bf 5}, 1629 (1972).

\bibitem{pg60} P. -G. de Gennes, Phys. Rev. {\bf 118}, 141 (1960).

\bibitem{jw97} J. C. Woicik, J. G. Pellegrino, B. Steiner, K. E. Miyano, S. G. Bompadre, L. B. Sorensen, T.-L. Lee, and S. Khalid,
Phys. Rev. Lett. {\bf 79}, 5026 (1997).

\bibitem{tobe} Changkun Xie, J. I. Budnick, B. O. Wells and J. C. Woicik, to be published.

\end{thebibliography}
\end{document}